# Electromagnetic Radiation Detection with Broad Spectral Sensitivity and Large Dynamic Range using Cantilever-based Photoacoustic Effect

Sucheta Sharma[1*], Toni Laurila[1], Jussi Rossi[2], Juho Uotila[3], Markku Vainio[2,4], Farshid Manoocheri[1] and Erkki Ikonen[1,5]

[1]*Metrology Research Institute, Aalto University, Espoo, Finland,*
[2]*Photonics Laboratory, Physics Unit, Tampere University, Tampere, Finland,*
[3] *Patria Aviation Oy, Tampere, Finland*
[4]*Department of Chemistry, University of Helsinki, Helsinki, Finland,*
[5]*VTT MIKES, Espoo, Finland*

A sensitive photoacoustic detection approach employing a silicon cantilever is investigated for power measurement of electromagnetic radiation. The technique which is actuated by pressure waves generated through radiation-induced heat, depicts high sensitivity for a considerably large spectral range from 325 nm to 1523 nm. The implemented method shows linear response in the measurement of radiation power from 15 nW to 6 mW, demonstrating a dynamic range of almost six orders of magnitude. A numerical model has been developed to analyse and optimise the measurement sensitivity using different dimensions of the cantilever which is one of the key components of the detection process. The numerical results are in good agreement with the experimentally obtained frequency response of the detection process. The power detection technique shows potential of finding future applications in the technologies that employ electromagnetic radiation detection for scientific studies and industrial purposes.

## I. INTRODUCTION

Photoacoustic (PA) effect is based on detecting pressure waves in a fluid medium, produced by the absorption of short-pulsed or modulated electromagnetic radiation [1-2]. Two key components, light absorbing medium and a pressure sensor [3-6], play major roles in such systems. Because of the constructional simplicity and detection sensitivity, the technique serves as a powerful tool for numerous applications in the field of high-resolution imaging, microscopy and in spectroscopic sensing of toxic, inflammable and explosive gases or particulate environmental pollutants, which pose major risk on health, safety and security [7-17]. For applications in the field of traceable measurement of radiation power, important prerequisites include a suitable radiation absorber material that responds to optical excitation along with a sensitive pressure detection technique that commonly employs electrical or optical methods [18-19]. The underlying mechanism is based on the production of pressure waves activated by radiation-induced heat generation in the optical absorber.

One of the common pressure detection methods in PA uses capacitive microphones. However, technologies with capacitive membrane microphones suffer from limitations such as the nonlinear response of the membrane in sensing the external pressure. Apart from that, the gap between the membrane and the backplane of the capacitive condenser cannot be reduced below a certain limit in order to improve the sensitivity, as the gas cannot anymore flow freely through such narrow region due to viscous effect [18]. The optical pressure sensing approach, on the other hand, is realised by monitoring the light beam deflected from elastic membranes or micro-mechanical silicon-cantilevers [19-20]. Cantilevers avoid some of the drawbacks of capacitive microphones, hence, they are used in many applications requiring very sensitive PA signal detection. The cantilever exhibits movement when pressure is applied due to the volume expansion of the carrier fluid. A compact interferometer is used for the detection process, which monitors the deflection of the cantilever tip to produce the PA signal [19]. The amplitude of the generated PA signal is directly proportional to the power of incident radiation. The technique works at room temperature and does not require cryogenic cooling.

*sucheta.sharma@aalto.fi

Most available semiconductor power detectors have spectral responsivity, which depends on the operating wavelength [21]. Pyroelectric detectors are excellent in sensing electromagnetic radiation over a wide spectral range with the noise floor for power measurement near 1 µW [22-23]. Golay cell is a well-known detection technology which uses the PA method. With proper window material, commercial solutions are available which are capable of covering a wide spectral range with the highest measurable power in the µW range [24].

In this work, we put forth studies on the potential of cantilever-based pressure sensor in PA detection of optical power. The proposed cantilever-based PA detection can be applied over the spectral range from UV to infrared. We have studied the frequency response and signal strength of two cantilevers of different dimensions by numerical simulations to optimise the detection sensitivity. With this method, the upper limit of detection can be made considerably high. Apart from that, to our knowledge, the analysis on optimising the detection sensitivity and improving the dynamic range of such systems for power measurement applications has not been reported earlier. Benefits of the proposed PA power measurement method include good linearity and broad spectral range. Here we have characterised the method for the wavelengths between 325 nm and 1523 nm. Measurements have been carried out in the range 15 nW - 6 mW, demonstrates the dynamic range of almost six orders of magnitude.

## II. EXPERIMENTAL SETUP AND METHODS

The schematic diagram of the experimental arrangement is depicted in FIG.1. The PA radiation-detection method is a combination of energy conversion processes: optical to thermal energy, which is followed by thermal to mechanical energy conversion. The experimental procedure is governed by three constitutive mechanisms: (i) Excitation using electromagnetic radiation; (ii) Temperature and pressure induced mechanical response; (iii) Optical detection of the mechanical response and generation of output PA signal. To actuate the mechanical response, the system requires such materials which help in efficient optical to thermal energy conversion. Studies have been carried out suggesting several black absorbing carbon-based materials such as carbon nanotubes, carbon black powders, carbon nanofibers [25-28], to name a few.

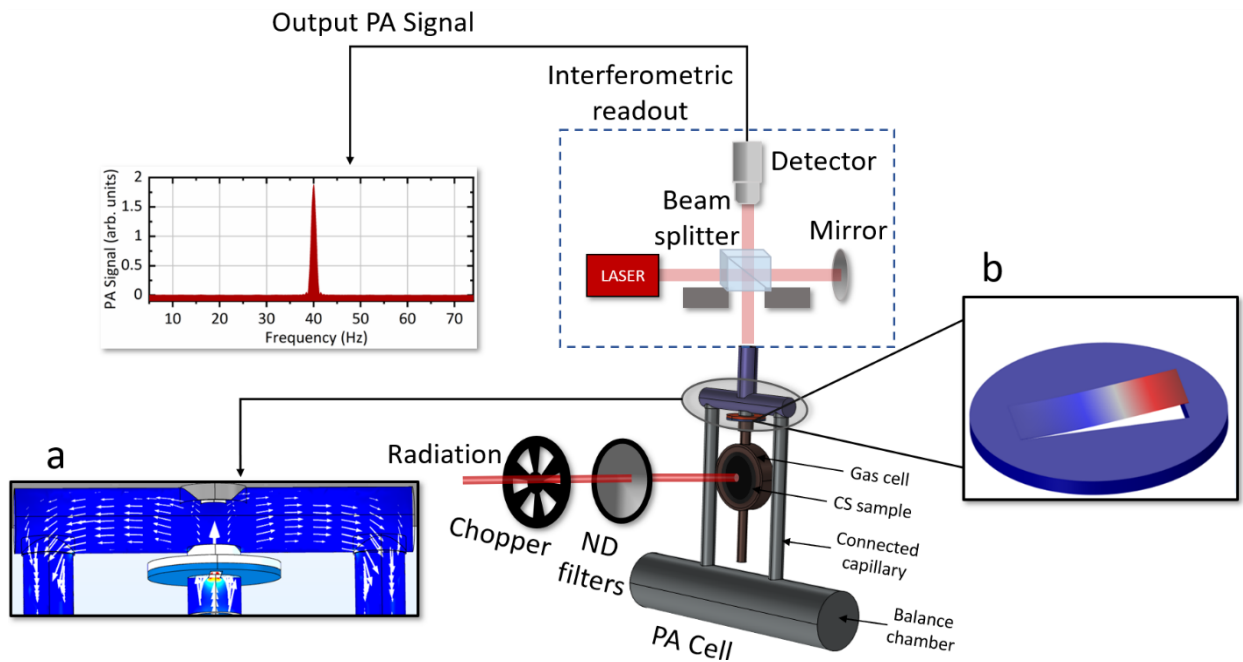

FIG.1. Schematic representation of the PA radiation detection system. It consists of three subsystems to receive input radiation in order to produce heat by using CS (candle soot) target, to produce mechanical response with the cantilever and, finally, to generate the PA signal from interferometric readout. There are two main gas domains: primary gas cell which directly heats up when the connected CS absorbs radiation and the balance chamber along with two capillaries that reduces the acceleration noise in the system. **a.** Velocity diagram of gaseous medium within the PA cell. The arrows represent how the induced pressure interacts with the cantilever to generate mechanical response. **b.** Cantilever displacement due to the generated pressure within the PA cell. The highest stress zone is concentrated at the tip of the cantilever.

*sucheta.sharma@aalto.fi

Candle soot (CS) is also a good candidate because of its cost-effective manufacturing process and strong PA response [29]. In the experimental setup the laser radiation first passes through a chopper, which produces intensity modulation of electromagnetic radiation that falls on the CS absorber. Here the frequency of the beam chopping was kept fixed at 40 Hz to achieve good signal-to-noise ratio. FIG.2 shows the response map of the CS absorber surface which was used in this study. The spatial scan of CS provides information about the region where the largest PA signal can be obtained. The experiment on spatial uniformity has been performed using 442 nm wavelength with a beam diameter of about 1.4 mm.

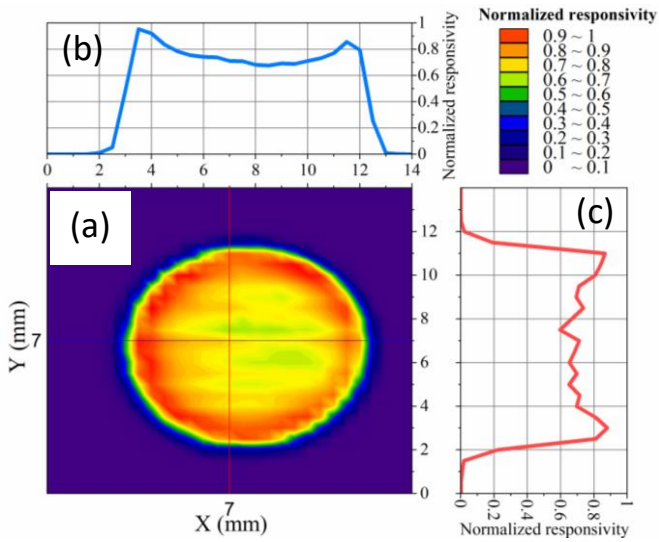

FIG.2. Spatial scan of the CS surface performed using 442 nm wavelength (a) The color distribution indicates the variation of the PA signal strength. (b) Normalized responsivity for the scanned region X= 0 – 14 mm and Y = 7 mm (indicated by horizontal line on the color map in (a)). (c) Normalized responsivity for the scanned region X = 7 mm and Y= 0 – 14 mm (indicated by vertical line in (a)).

The output PA signal as shown in the diagram of FIG.1 represents the response of the cantilever which, for this particular example, exhibited displacement at 40 Hz movement frequency in response to the intensity variation of the radiation source at the 40 Hz chopping frequency. By using a variable neutral density (ND) filter, the power of the input radiation can be attenuated to study the detector's response to varying power levels.

The amplitude of PA signal for a given incident radiation can be expressed as: $S_{PA} = \sigma_{PA} P$ where $\sigma_{PA}$ is the detection sensitivity and $P$ is the power of incident radiation. Sensitivity $\sigma_{PA}$ depends on the properties of the light absorber, the PA cell and the cantilever. Thus, the cantilever has a significant role in improving the PA system response. In order to improve $\sigma_{PA}$, we have focused on the cantilever design by mainly analyzing the suitable dimensions. In addition to the deflection amplitude of the cantilever, the resonance frequency is also an important factor for achieving overall better system performance. In order to have system stability where small changes in modulation frequency do not significantly change the PA amplitude, the operating frequency is generally chosen lower than the resonance frequency of the system. Hence, a proper design of the cantilever is crucial which exhibits higher deflection for improving sensitivity and also the information of the resonance frequency helps in determining the operating frequency region of the system.

### III. NUMERICAL MODEL

We have developed a numerical model of the experimental setup using COMSOL Multiphysics® v. 5.4.0.388. The PA cell with gas cell radius of 6.34 mm and height of 3.28 mm, contains the gaseous medium, e.g., air or helium (He), under a reference pressure which is modified when the medium is exposed to periodic temperature variation. The resulting periodically varying pressure waves initiate the cantilever movement to generate the PA signal. The cantilever displacement is computed at different frequencies to study the frequency response of the system. The frequency response of the cantilever-based PA detection gives important information about how the sensitivity can be improved.

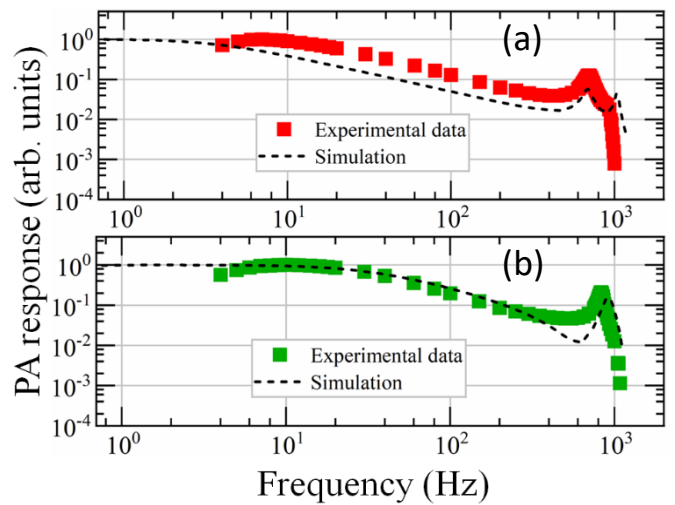

FIG.3. Frequency response study of cantilever-I for (a) air and (b) He. The experimental data represent the PA response at different chopping frequencies. The dotted line depicts the result of numerical simulation which shows the amplitude of cantilever displacement for different values of modulating frequencies.

*sucheta.sharma@aalto.fi

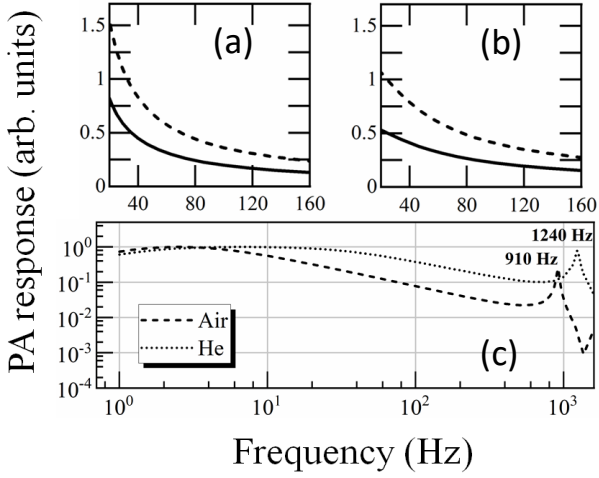

FIG.4. Result of numerical simulation for (a) air and (b) He on the improvement of signal strength in the frequency region from 20 Hz to 160 Hz for cantilever-I (dashed line) compared to cantilever-II (solid line). (c) Frequency response of cantilever-II. The resonance frequency becomes higher when the system operates with He as gaseous medium compared to air.

For small perturbations around the steady background values, the dependent variables take the form,

$$p = p_b + p'e^{j\omega t} \qquad (1)$$
$$\mathbf{u} = \mathbf{u_b} + \mathbf{u'}e^{j\omega t} \qquad (2)$$
$$T = T_b + T'e^{j\omega t} \qquad (3)$$
$$\rho = \rho_b + \rho'e^{j\omega t} \qquad (4)$$

where $p$ is the pressure, $\mathbf{u}$ is velocity field, $T$ is temperature, $\rho$ fluid density and $\omega$ is angular frequency. The background pressure, velocity, temperature and fluid density are $p_b$, $\mathbf{u_b}$, $T_b$ and $\rho_b$, respectively. It can be assumed that there is no contribution of background velocity, thus $\mathbf{u_b} = 0$. The values of $T_b$ and $p_b$ were chosen to be 293 K and 1013.25 hPa, respectively. The governing equations are momentum equation, continuity equation and energy equation. In frequency domain, the equations are expressed as,

$$j\omega\rho_b \mathbf{u} = \nabla \cdot \left[-p\mathbf{I} + \mu(\nabla \mathbf{u} + (\nabla \mathbf{u})^T) - ((2/3\,\mu) - \mu_B)(\nabla \cdot \mathbf{u})\mathbf{I}\right] \qquad (5)$$

$$j\omega\rho + \nabla \cdot (\rho_b \mathbf{u}) = 0 \qquad (6)$$

$$\rho_b C_p (j\omega T + \mathbf{u} \cdot \nabla T_b) - T_b \alpha_p (j\omega p + \mathbf{u} \cdot \nabla p_b) = -\nabla \cdot (-K\nabla T) + Q \qquad (7)$$

The parameters $\mu$, $\mu_B$, $C_p$, $\mathbf{I}$, $Q$ and $K$ are the dynamic viscosity, bulk viscosity, heat capacity at constant pressure, unit tensor, heat source and thermal conductivity, respectively [30]. In addition to the above equations, there is the linearized equation of state, $\rho = \rho_b(\beta_T p - \alpha_p T)$ where $\alpha_p$ is the coefficient of thermal expansion and $\beta_T$ is the isothermal compressibility. In the case of air and He as the gaseous media, the expressions for $\alpha_p$ and $\beta_T$ can be considered under the case of ideal gas approximation where the value of $\alpha_p$ is the inverse of $T_b$ and $\beta_T$ is the inverse of $p_b$.

TABLE I summarizes the input parameters of the model. The remaining input parameters include the density $\rho_{Si}$, Young's modulus $E$ and Poisson's ratio $\nu$ of the cantilever material silicon, which are 2329 kg/m³, 170×10⁹ Pa and 0.28, respectively. The thickness of the viscous boundary layer is given by $\delta_{visc} = \sqrt{2\mu/\omega\rho_b}$. The layer thicknesses for air and He at 40 Hz, 20 °C and 1013.25 hPa are thus 0.088 mm and 0.246 mm, respectively. The value of Prandtl number $(P_r)$, $P_r = \delta_{visc}^2/\delta_{therm}^2$, is 0.7 for air and 0.69 for He, where $\delta_{therm} = \sqrt{2k/\omega\rho_b C_p}$ is the thermal boundary layer thickness [31-32].

TABLE I. Constituting material parameters of the numerical model.

| Model parameters | Air | He |
| --- | --- | --- |
| $K$ (W/mK) | 0.02565 | 0.14929 |
| $C_p$ (J/kg K) | 1015.12 | 5196.49 |
| $\rho_b$ (kg/m³) | 1.2032 | 0.1663 |
| $\mu$ (Pa s) | 1.817×10⁻⁵ | 1.973×10⁻⁵ |
| $\mu_B$ (kg s) | 1.81×10⁻⁵ | 1.96×10⁻⁵ |

TABLE II. Comparison between experiment and simulation for calculating the resonance frequency ($f_0$) of the PA detection system with cantilever-II.

| Material | $f_0|_{Sim.}$ in Hz | $f_0|_{Expt.}$ in Hz | Relative difference $\dfrac{(f_0|_{Expt.} - f_0|_{Sim.})}{f_0|_{Sim.}} \times 100\,\%$ |
| --- | --- | --- | --- |
| Air | 690 | 700 | + 1.5 % |
| He | 920 | 830 | - 9.8 % |

*sucheta.sharma@aalto.fi

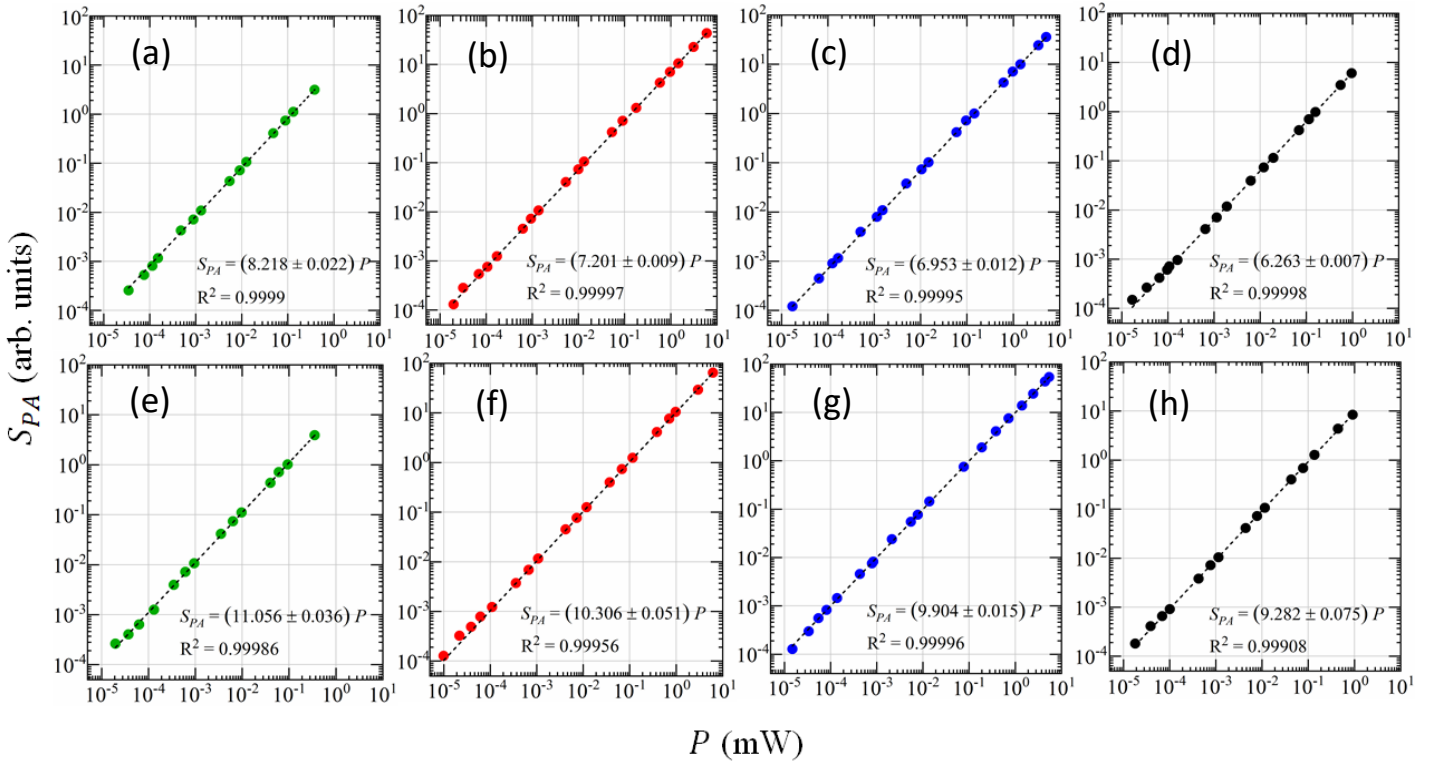

FIG.5. Generated PA response with incident radiation power from ~ 15 nW to ~ 6 mW at 40 Hz chopping frequency under gaseous medium air (row 1: (a) - (d)) and He (row 2: (e) - (h)). The excitation wavelengths are: 325 nm (column 1), 442nm (column 2), 633nm (column 3) and 1523 nm (column 4). The dashed lines represent the linear fit (all linear regressions are passing through the origin). The values of $R^2$ for all the datasets are close to 1, suggesting good agreement between the linear fit and the experimental data points. The PA signal amplitude $(S_{PA})$, which varies linearly with optical power $(P)$, higher detection sensitivity $(\sigma_{PA})$ for He compared to air for the wavelengths between 325 nm and 1523 nm.

To validate the model, the frequency response of the system for a cantilever (cantilever-I) having dimensions 6 mm × 0.7 mm × 5 μm (length × width × thickness: $l \times w \times t$) has been first compared with experimental data. TABLE II and FIG.3 show the comparison between the experimentally and numerically obtained data on the resonance frequency of the system. As shown in FIG.3 (a) & (b), the model is able to capture the increase in the resonance frequency when the medium is changed from air to He. This effect is also visible in the experimental data. The result demonstrates how the frequency response changes with change of the carrier gas.

To analyze the effect of cantilever dimensions for improving the sensitivity, another cantilever (cantilever-II) with $l \times w \times t$ : 4 mm × 1 mm × 10 μm has been checked by using the model. The frequency response of cantilever-II is shown in FIG.4 (c). The resonance frequency, in this case, is even higher compared to cantilever-I but the deflection amplitude, as shown in FIG.4 (a) & (b), is less compared to cantilever-I for air and He, respectively, at 40 Hz chopping frequency. However, both cantilevers have considerably high resonance frequencies compared to the operational frequency, which in our case, typically stays within 30 - 50 Hz. Hence, the experiments have been carried out on cantilever-I which has higher deflection amplitude and the following sections contain the results on the performance of the PA power detector with improved sensitivity using cantilever-I.

## IV. RESULTS AND DISCUSSION

FIG.5 (a)-(h) show linearity of the PA peak amplitude with respect to incident power. The experimental data indicate the lowest detectable power of about 15 nW. Another important feature of this detection process is the large dynamic range. FIG.5 (b), (c), (f) and (g) show that the detection technique can be applied from 15 nW to 6 mW to achieve a dynamic range of nearly six orders of magnitude in both air and He.

The evolution of $S_{PA}$ with the increments of the incident radiation power of 1523 nm wavelength for gaseous medium air and He, is shown in FIG. 6 (a)-(b). The gray area in the graph represents the background noise which limits the lowest detectable power. The wavelength dependence of the PA detection sensitivity, $\sigma_{PA}$ is shown in FIG.7 (a) and (b) for 325 nm, 442 nm, 633 nm and 1523 nm incident radiation wavelengths. The results indicate that the detection process has almost similar values of $\sigma_{PA}$ which varied from $(8.218 \pm 0.022)$ to $(6.263 \pm 0.007)$ arb. unit /mW for air and from $(11.056 \pm 0.036)$ to $(9.282 \pm 0.075)$ arb. unit/mW for He, within the spectral range of 325 nm – 1523 nm. The PA signal arb. units are the same for all figures.

*sucheta.sharma@aalto.fi

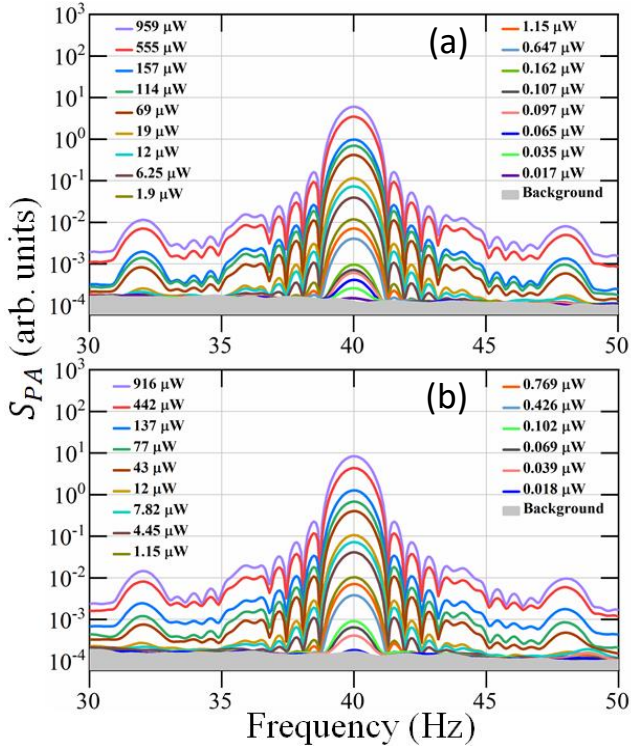

FIG.6. Photoacoustic peak amplitude increment with the increase of incident radiation intensity for gaseous medium (a) air and (b) He. The grey region represents the background signal.

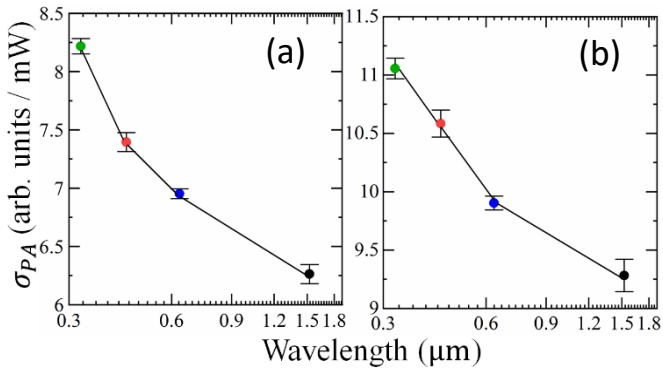

FIG.7. Detection sensitivities, $\sigma_{PA}$, as measured from the slope of the linear fit in FIG. 5, for different wavelengths of incident radiation. The value of the estimated $\sigma_{PA}$ decreases by 15 to 25 % as the wavelength increases for both (a) air and (b) He. The uncertainty bars describe the combined standard uncertainty.

Apart from the uncertainty due to the linear fitting, the estimated sensitivities of FIG.7 have measurement uncertainties due to the reference detector calibration and the spatial nonuniformity of the PA responsivity.

*sucheta.sharma@aalto.fi

Silicon photodiode detectors traceable to a pyroelectric detector [33] and to a predictable quantum efficient detector [34] are used as the reference detectors for the 325 nm and 442-633 nm wavelengths, respectively, whereas an InGaAs detector traceable to the pyroelectric detector [35] is used as the reference detector at 1523 nm. The reference power measurement has a standard uncertainty of 0.6 % at 325 nm, 0.25 % at both 442 nm and 633 nm and 1.15% at 1523 nm.

The nonuniform spatial responsivity as indicated in FIG.2 causes variations in the PA sensitivity estimation because of differences in the beam diameter at the used wavelengths and repeatability of the measured sensitivities. The $1/e^2$ beam diameter is from 0.8 mm to 1.0 mm at other wavelengths except at 442 nm where it is 1.4 mm. A correction factor of $1.027 \pm 0.010$ was determined for the value of $\sigma_{PA}$ measured at the wavelength of 442 nm to be compatible with the data at other wavelengths. The standard uncertainty due to spatial nonuniformity of the responsivity and repeatability of the measurements is 1.0 % at 442 nm and 0.5 % at other wavelengths. Adding quadratically the different uncertainty components of $\sigma_{PA}$ gives the combined relative standard uncertainties of 0.8 % at 325 nm, 1.1 % at 442 nm, 0.6 % at 633 nm, and 1.3 % to 1.5 % at 1523 nm. At 1523 nm, the lower uncertainty is for air-filled and the higher for He-filled PA cell, whereas at other wavelengths there is no considerable amount of difference observed in the total uncertainty for air and He.

## V. CONCLUSION

In conclusion, we implemented a cantilever-based PA radiation detection system, which is capable of detecting radiation power level ranging from 15 nW to ~ 6 mW, indicating a linear dynamic range of nearly six orders of magnitude. We have developed a numerical model that can be employed to optimize the PA detector design for radiation power measurements. Two different cantilever dimensions have been studied through simulations. Cantilever-I which showed better signal strength compared to cantilever-II, has been employed for the experiments. The difference between the experimental result and the predicted resonance frequency by the model is within 10 %. We have verified that the power detection scheme can be employed for the spectral region ranging from 325 nm to 1523 nm. To check the linearity and the detection sensitivity, we performed experiments with four laser sources of wavelengths 325 nm, 442 nm, 633 nm and 1523 nm where the power of incident radiation was extended up to nearly 6 mW from 15 nW by using ND filters. The result showed good linearity for both air and He. In the future, the detection ability will be tested up to THz range after employing a suitable window material in the setup. However, the present work verifies the potential of the cantilever-based PA detection system which can replace the need of multiple detection schemes for measuring electromagnetic radiation power from 15 nW to 6 mW with spectral range coverage from 325 - 1523 nm.

## ACKNOWLEDGEMENTS

We thank Janne Askola for helping with the experiments on the spatial scan. This project is funded by the Academy of Finland Flagship Programme, Photonics Research and Innovation (PREIN), decision number: 320167. The project is also a part of Universal electromagnetic radiation detector (UNIDET) research project, which is funded by the Academy of Finland (Project numbers 314363 and 314364).

*sucheta.sharma@aalto.fi*

*sucheta.sharma@aalto.fi*